\documentclass[twocolumn,aps,superscriptaddress,prl]{revtex4}
\usepackage{palatino}
\usepackage[latin1]{inputenc}
\usepackage{epsf}
\usepackage{amsmath,amssymb}
\usepackage{latexsym}
\usepackage{calc}
\usepackage{xcolor}
\usepackage{listings}

\usepackage[normalem]{ulem}

\usepackage{graphicx}
\usepackage{hyperref}
\newcommand{\ben}{\begin{equation}}
\newcommand{\een}{\end{equation}}
\newcommand{\bea}{\begin{eqnarray}}
\newcommand{\eea}{\end{eqnarray}}

\def\sss{\scriptscriptstyle\rm}

\def\1s{_{1,\sss S}}
\def\2s{_{2,\sss S}}

\def\s{_{\sss S}}
\def\xc{_{\sss XC}}

\def\Hxc{_{\sss HXC}}

\def\bigsss{\displaystyle\rm}  
\def\a{_{\bigsss a}}

\def\Hxcqqp{_{{\sss HXC}\,qq'}}

\def\xcqqp{_{{\sss XC}\,qq'}}

\def\xcQQ{_{{\sss XC}\,QQ}}
\def\xcQQp{_{{\sss XC}\,QQ'}}

\def\br{{\bf r}}

\def\dulR{{\underline{\underline{\bf R}}}}

\usepackage{listings}
\usepackage{xcolor}

\begin{document}
 \title{Capturing the Elusive Curve-Crossing in Low-Lying States of Butadiene With Dressed TDDFT  
}
 \author{Davood B. Dar}
 \affiliation{Department of Physics, Rutgers University, Newark 07102, New Jersey USA}

 \author{Neepa T. Maitra}
 \affiliation{Department of Physics, Rutgers University, Newark 07102, New Jersey USA}
 \email{neepa.maitra@rutgers.edu}
 \date{\today}

\begin{abstract}
  A striking example of the need to accurately capture states of double-excitation character in molecules is seen in predicting photo-induced dynamics in small polyenes. Due to the coupling of electronic and nuclear motions, the dark 2$^1$Ag state, known to have double-excitation character, can be reached after an initial photo-excitation to the bright $^1$Bu state via crossings of their potential energy surfaces. However, the shapes of the surfaces are so poorly captured by most electronic structure methods, that the crossing is missed or substantially mis-located. We demonstrate that the frequency-dependent kernel of dressed TDDFT beyond Tamm-Dancoff successfully captures the curve-crossing, providing an energy surface close to the highly accurate but more expensive $\delta$-CR-EOMCC(2,3) benchmark reference.  This, along with its accurate prediction of the excitation character of the state makes dressed TDDFT a practical and accurate route to electronic structure quantities needed in modeling ultrafast dynamics in molecules.
\end{abstract}
\maketitle
\section{Introduction}
Linear polyenes have long been the focus of intensive experimental and theoretical work owing to their importance in biological and  technologically important systems~\cite{HK72,SK72,SOK76,HK78,CD87,OZZ91, FLMSST98, CBBGF04,LM07,LM09, KOK13, B22}. In particular the photophysical behavior of the smallest polyenes, \textit{s-trans}-butadiene and \textit{s-trans}-hexatriene, is of significant interest in organic electronics and photovoltaics~\cite{PKSCG02}, and are viewed as representative motifs of polyenes responsible for triggering vision~\cite{OZZ91}. 
Despite this longstanding interest, accurately describing the electronic structure and photodynamics of linear polyenes has proven challenging for theoretical methods, leading to 
unreliable predictions of relaxation processes through conical intersections after photo-excitation.
These difficulties stem from the nature of the low-lying excited states in these molecules, particularly the dark 2$^1$Ag state, due to its partial  doubly-excited character~\cite{OKS78,LC02,KMLJ24}, which many single reference methods struggle to capture. 
Conventional computational techniques are often unreliable for these excited state: it is well-documented that equation of motion coupled cluster with singles and doubles (EOM-CCSD) falls short~\cite{LBSCJ19,PSLPF21}, while complete active space self-consistent field methods (CASSCF and CASPT2) need a careful evaluation of the active space~\cite{LM09,PSLPF21}.
Even though these approximate methods can be adjusted to describe a single-point calculation of  2$^1$Ag accurately, their reliability across geometry variations is suspect. 
They typically incorrectly predict the crossing of the potential energy surfaces of the two lowest excited states, 1$^1$Bu and 2$^1$Ag~\cite{PSLPF21}. As a result, many studies have struggled to describe the ultrafast dynamics of polyenes with the precision required for predicting observed results.

While it is known that the challenge essentially lies in incorporating double excitations in the descriptions of the excited states, most computational techniques that do accurately handle these states tend to become very expensive for longer polyenes. For instance, the $\delta$-CR-EOMCC(2,3) approach of Refs.\cite{FLWPW11,PHA15}, which corrects the excited-state potentials obtained with EOMCCSD for the effects of triple excitations, correctly captures the curve crossing in s-trans-butadiene and s-trans-hexatriene~\cite{PSLPF21}. However, developing a method that achieves comparable accuracy at a lower computational cost would significantly benefit calculations for larger systems.

In principle, time-dependent density functional theory (TDDFT)~\cite{RG84,C95,PGG96} could serve as a low-cost alternative due to its famously tolerable scaling. However, in practice the adiabatic approximation inherent in its standard linear response formulation ignores double excitation contributions, and as a result it struggles to accurately describe the energy ordering and shapes of the surfaces for these states~\cite{PSLPF21}. Other DFT-based methods that have been explored for double-excitations include orbital-optimized $\Delta$SCF~\cite{HH21}, ensemble-DFT~\cite{YPBU17,CF24,GKP21}, and pp-RPA~\cite{YLZY24}. 
Within TDDFT, to go beyond adiabatic TDDFT (ATDDFT) so-called dressed TDDFT approaches have been developed~\cite{TAHR99,TH00,CZMB04,C05,MW09,MMWA11,RSBS09,SROM11,HIRC11,CH12,CH16,M22,DM23}, in which a frequency-dependent kernel is designed to incorporate contributions from double excitations. The original version operated within the Tamm-Dancoff approximation~\cite{MZCB04,CZMB04,C05,MW09,CH16}, denoted here as dressed Tamm-Dancoff (DTDA), and 
has successfully predicted the energies of double excitations across a variety of molecules~\cite{CZMB04,HIRC11,MW09,MMWA11,M22}. However, because the Tamm-Dancoff approximation inherently does not preserve oscillator strengths, DTDA cannot be used reliably  for transition properties including the non-adiabatic couplings between states needed for dynamics. In Ref.~\cite{DM23}, we derived a frequency-dependent kernel designed to function within the full TDDFT linear response framework, ensuring that the oscillator strength sum rule is preserved. Its performance on model as well as real systems, including the lowest $^1D$ excitations of the Be atom and LiH molecule across varying interatomic distances was found to be satisfactory. This kernel redistributes the transition density of a Kohn-Sham (KS) single excitation into a mix of single and double excitations, resulting in better predictions for transitions between excited states, as demonstrated in Ref.~\cite{DM23}.
The examples in Ref.~\cite{DM23} involved cases where one single excitation couples with a double excitation  while more commonly, and in the case of  the $2^1$Ag state of linear polyenes, the state is composed of several single-excitations mixing with a double-excitation with respect to the ground-state KS reference. Here we extend the approach of Ref.~\cite{DM23} to scenarios where multiple single excitations couple with a double excitation. 
We show that the potential energy surfaces of 1$^1$Bu and 2$^1$Ag for \textit{s-trans}-butadiene resulting from our dressed TDDFT (DTDDFT) 
display a crossing close to that predicted by the highly-accurate $\delta$-CR-EOMCC(2,3)~\cite{PHA15, PSLPF21} method, which we use as a benchmark in this study.

\section{Theory}
The fundamental of idea of DTDDFT  beyond DTDA  was presented  in Ref.~\cite{DM23}, motivated by needing accurate oscillator strengths and transition densities, in addition to the energies. The equations were derived there for the case when the state of interest involves one single excitation out of the KS ground-state reference mixing with one double excitation. One begins  with the time-independent Schrodinger equation in the following form:
\begin{equation}
(H - E_0)^2 \Psi = \omega^2 \Psi
\end{equation}
where $E_0$ is the ground state energy and $\omega$ represents the excitation frequency. In the truncated Hilbert space of the doubly-excited determinant $|D\rangle$ and singly-excited determinant $|Q\rangle$, diagonalization of the operator $(H-E_0)^2$ leads to 
\bea
    \omega^2&=&(H_{QQ} - E_0)^2 + \vert H_{QD} \vert^2 \nonumber\\ 
    &\times& \Bigg[ 1 +\frac{(H_{QQ} + H_{DD} - 2E_0)^2}
               {\omega^2 - \left[ (H_{DD} - E_0)^2 + \vert H_{QD} \vert^2 \right]} 
               \Bigg]
\label{eq:diag}
\eea
where $H_{AB}$ are the matrix elements of the true Hamiltonian between KS states.  Obtaining excitation energies in TDDFT however does not proceed through diagonalization, but instead through a linear response procedure, where the KS single excitations get corrected to the exact ones via the exchange-correlation (xc) kernel $f\xc(\br,\br',\omega)$~\cite{C95,PGG96}. Ref.~\cite{DM23} mapped Eq.~(\ref{eq:diag}) on to the TDDFT linear response equation truncated to one single excitation, and thereby extracted a frequency-dependent xc kernel. The first term in Eq.~\ref{eq:diag}, which is the contribution to $\omega^2$ from the single excitation, was replaced with the ATDDFT small matrix approximation (SMA) value, while the residual part,  with $E_{0}$ replaced by $H_{00}$ to balance the errors in accordance with what was done in Ref.~\cite{MZCB04},  was interpreted as a frequency-dependent kernel in a dressed SMA (DSMA):
\bea
\label{eq:DSMA_0}
f\xcQQ^{\rm DSMA_0} (\omega) &=& f\xcQQ^{\rm adia} + \frac{\vert H_{QD}\vert^2}{4\nu_q}\\
\nonumber
&\times& \left(1 + \frac{(H_{QQ}+H_{DD}-2H_{00})^{2}}{\left[\omega^{2}-\left((H_{DD}-H_{00})^{2}+H_{QD}^{2}\right)\right]}\right)\,.
\eea
Here $f\xcQQ^{\rm adia}$ denotes the diagonal matrix element of an adiabatic xc-kernel, whose general expression is 
\begin{equation}
\small
    f^{\rm adia}\xcQQp(\omega)=\int d^3r \int d^3r' \, \Phi_Q^*(\mathbf{r}) \, f^{\rm adia}\xc(\mathbf{r}, \mathbf{r}', \omega) \, \Phi_{Q'}(\mathbf{r'})
    \label{eq:fxcmatrixelt}
\end{equation}
in which $\Phi_Q^*(\mathbf{r})=\phi_i(\br)\phi_a(\br)$ is the product of orbitals involved in the KS single excitation $Q:i\rightarrow a$. Eq.~(\ref{eq:DSMA_0}) represents a dressed SMA  (DSMA), and this kernel should be used within the SMA. Further, two variants  of Eq.~(\ref{eq:DSMA_0}) were proposed by making the following replacements 
\bea
    (H_{QQ} + H_{DD} - 2H_{00})^2&\xrightarrow {\rm{DSMA}\s}& (\nu_Q+\nu_D)^2\nonumber\\
    (H_{DD} - H_{00})^2 &\xrightarrow { \hspace{1cm}  } &\nu_D^2
    \label{rep_s}
\eea
and
\bea
    (H_{QQ} + H_{DD} - 2H_{00})^2&\xrightarrow{\rm{DSMA}\a}& (\Omega_{Q}^A+\Omega_{s1}^A+\Omega_{s2}^A)^{2}\nonumber\\
    (H_{DD} - H_{00})^2 &\xrightarrow { \hspace{1cm}  } &(\Omega_{s1}^A+\Omega_{s2}^A)^{2}
    \label{rep_a}
\eea
where $\nu_Q$ and $\nu_D \equiv \nu_{s1} + \nu_{s2}$ are the KS frequencies of the single and double excitations, obtained from KS orbital energy differences (e.g. $\nu_Q = \epsilon_a - \epsilon_i$) and $\Omega_{Q (s1, s2)}^A$ are the ATDDFT-corrected frequencies of the single excitation $Q$ and the two single excitations $s_1, s_2$ that compose the KS double-excitation. The s subscript on DSMA$\s$ denotes the replacement of the diagonal Hamiltonian matrix elements with KS energies, while the a subscript on DSMA$\a$ indicates replacement by ATDDFT~\cite{DM23}.
These two variants are much easier  to implement computationally than DSMA$_0$ as they require fewer two-electron matrix elements to be computed: Apart from $H_{QD}$ one can extract all the other quantities directly from the  ATDDFT calculation.

To extend this approach to cases where more than one single excitation couples with a double excitation, we follow a similar strategy as above but explicitly mix in more than one single KS excitation by first expressing the single-excitation component $|Q\rangle$ of the interacting state  as a linear combination of all the KS single excitations $|q\rangle$
 \begin{equation}
     |Q\rangle = \displaystyle\sum_{q} c_q |q\rangle.
     \label{linear_comb_sing}
 \end{equation}
Diagonalization in the space of all these single excitations plus the double excitation that couples with them again leads to Eq.~\ref{eq:diag} with the following redefinitions 
\begin{equation}
H_{QQ} = \sum_{q,q'} c_{q}^* c_{q'} H_{qq'} 
\end{equation}
and 
\begin{equation}
H_{QD} = \sum_{q} c_q^* H_{qD}\,.
\end{equation}
Using these forms, Eq.~(\ref{eq:diag}) becomes 
\bea
\omega^2 = \left(\sum_{q,q'} c_q^* c_{q'} H_{qq'}- H_{00})\right)^2+ \sum_{q,q'} c_q^* c_{q'}  H_{qD} H_{Dq'}\nonumber\\
\times\left[ 1 + \frac{(\sum_{q,q'} c_q^* c_{q'} H_{qq'} + H_{DD} - 2H_{00})^2}{\omega^2 - \left[ (H_{DD} - H_{00})^2+ \sum_{q,q'} c_q^* c_{q'}  H_{qD} H_{Dq'} \right]} \right].
\label{eq:diag2}
\eea
As in the case of a single single-excitation coupling to a double, this result of diagonalization is used to inspire an approximation for a frequency-dependent TDDFT xc kernel for the present case of several single-excitations coupling to a double. This time, we are beyond the SMA, and consider the full TDDFT matrix,
\begin{equation}
    \Omega(\omega)_{q q'} = \nu^2_{q} \delta_{q q'} + 4 \sqrt{\nu_{q}\nu_{q'}} f\Hxcqqp(\omega)\,
    \label{eq:casida_eq}
\end{equation}
whose eigenvalues give the (in-principle exact) squares of the frequencies $\omega^2$ of the true system.
Here $\nu_{q}$ is the frequency of the KS single excitation $\vert q \rangle$ and $f\Hxcqqp(\omega)$ is matrix element (Eq.~(\ref{eq:fxcmatrixelt})) of the Hartree-xc kernel
\begin{equation}
    f\Hxc(\mathbf{r}, \mathbf{r}', \omega) = \frac{1}{|\mathbf{r} - \mathbf{r}'|} + f\xc(\mathbf{r}, \mathbf{r}', \omega)\,.
\end{equation}
Now, to construct the approximation to the xc kernel, we first require that in the limit of all the couplings between the single and double excitations in Eq.~(\ref{eq:diag2}) going to zero, $H_{qD}\rightarrow0$, the procedure reduces to Eq.~(\ref{eq:casida_eq}) with an adiabatic approximation to the kernel. The justification is that ATDDFT is typically accurate for states of single-excitation character, provided the ground-state approximation used has appropriate long-ranged characters when needed~\cite{SKB09,M16,M17,K17b}.  That is, we replace the diagonalization amongst the KS singles in Eq.~(\ref{eq:diag2}) with the adiabatic part of Eq.~(\ref{eq:casida_eq}) i.e.
\begin{equation}
\left(\sum_{q,q'} c_q^* c_{q'} H_{qq'}- H_{00})\right)^2 \Rightarrow \nu^2_{q} \delta_{qq'} + 4 \sqrt{\nu_{q}\nu_{q'}} f^{\rm{adia}}\Hxcqqp\,.
\end{equation} 
This then suggests the following approximation for the frequency-dependent DTDDFT xc kernel matrix:
\bea
f^{\mathrm{DTDDFT_0}}\xcqqp(\omega)&=&f^{\rm{adia}}\xcqqp + X^0_{qq'}(\omega)
\label{kernel_0}
\eea
where 
\bea
X^0_{qq'}(\omega)=\frac{H_{qD}H_{Dq'} }{4 \sqrt{\nu_{q}\nu_{q'}}}\left[ 1 + \frac{(H_{qq'}+ H_{DD} - 2H_{00})^2}{\omega^2 -  (H_{DD} - H_{00})^2} \right].
\label{dressing_0}
\eea
We dropped the last term in the denominator of Eq.~(\ref{eq:diag2}) for the following two reasons. First, it would give the kernel an unphysical dependence on the choice of arbitrary sign for the states $q,q'$ (that is, if the arbitrary sign of the state $q$ was chosen flipped, all matrix elements of $\Omega_{qq'}$ should simply flip in order to yield the same energies). Second, taking the Tamm-Dancoff limit of our approximation, where ``backward" transitions are neglected, it should reduce to the previously-derived DTDA of Refs.~\cite{CZMB04,MW09}, but this only happens if this term is dropped (see the Supporting Information). 

Eq.~(\ref{kernel_0}), together with two variants that reduce the computational cost, analogous to Eqs.~(\ref{rep_s})--(\ref{rep_a}), 
\bea
    f\Hxcqqp^{\mathrm{DTDDFT\s}}(\omega)=f^{\rm{adia}}\Hxcqqp + X^{\s}_{qq'}(\omega)
   \label{kernel_s}\\
f\Hxcqqp^{\mathrm{DTDDFT\a}}(\omega)=f^{\rm{adia}}\Hxcqqp + X^{\a}_{qq'}(\omega)
 \label{kernel_a}
\eea
where
\bea
X^{\s}_{qq'}(\omega)=\frac{H_{qD}H_{Dq'} }{4 \sqrt{\nu_{q}\nu_{q'}}}\left[ 1 + \frac{(\nu_{q}+\nu_D)(\nu_{q'}+\nu_D)}{\omega^2 - \nu_D^2} \right]
 \label{dressing_s}
\eea
and 
\bea
X^{\a}_{qq'}(\omega)=\frac{H_{qD}H_{Dq'} }{4 \sqrt{\nu_{q}\nu_{q'}}} \left[ 1 + \frac{(\Omega_{q}^A+\Omega_{s1}^A+\Omega_{s2}^A)(\Omega_{q'}^A+\Omega_{s1}^A+\Omega_{s2}^A)}{\omega^2 - (\Omega_{s1}^A+\Omega_{s2}^A)^{2}} \right]
\label{dressing_a}
\eea
form the central equations of this paper. They define three non-adiabatic kernel approximations that account for a double-excitation mixing with multiple single excitations. On the one hand, these equations extend the DSMA approximations of Ref.~\cite{DM23} to the case where the state of double-excitation character involves several KS single excitations instead of just one, and on the other hand, they extend the approximation of Refs.~\cite{CZMB04,MW09,C05} to include the ``backward excitations" that are needed to restore the oscillator strength sum-rule. As stated above, for computational simplicity reasons, we will utilize $f^{\text{DTDDFT}\s}\xc$ and $f^{\text{DTDDFT}\a}\xc$.

For practical application, the kernel needs to be integrated into the established methodology for computing excitations from TDDFT. This typically proceeds via diagonalizing the pseudo-eigenvalue equation
\begin{equation}
    \begin{pmatrix}
A  & B  \\
B^* & A^*
\end{pmatrix}
\begin{pmatrix}
X \\
Y
\end{pmatrix}
=
\omega^2
\begin{pmatrix}
-1 & 0 \\
0 & 1
\end{pmatrix}
\begin{pmatrix}
X \\
Y
\end{pmatrix}
\label{eq:casida_matrix}
\end{equation}
where $ A_{qq'} = \delta_{qq'} \nu_q+2f^{\text{adia}}\Hxcqqp$ and $B_{qq'} = 2f\Hxcqqp^{\text{adia}}$. Assuming real orbitals, this is equivalent to diagonalizing the matrix~\cite{C95,c96,DH05}
\begin{equation}
    \Omega = (A - B)^{1/2} (A + B ) (A - B)^{1/2}.
    \label{eq:alternate_casida}
\end{equation}
Implementing the dressed kernels Eqs.~(\ref{kernel_0})--(\ref{kernel_a}) amounts to adding one of the dressing terms defined in Eqs.~(\ref{dressing_0}),(\ref{dressing_s} and (\ref{dressing_a}) to matrices A and B.
The oscillator strengths for the excited states are computed from the eigenvectors, $G_I$ of Eq.\ref{eq:casida_matrix} normalized according to
\bea
    G_{I}^{\dag}\left(1-\left[\frac{\partial\Omega(\omega)}{\partial\omega^{2}}\right]_{\omega=\omega_{I}}\right)G_{I}=1
    \label{normalization}
\eea

\subsection{Implementation}
\label{sec:Implementation}
Before applying DTDDFT to a particular molecular system, we first outline the computational process for implementing DTDDFT. 
As in previous dressed approaches~\cite{MZCB04,CZMB04,C05,MW09,MMWA11,HIRC11,DM23}, the dressings in  Eq.~(\ref{dressing_0}),(\ref{dressing_s} and (\ref{dressing_a}) operate  as \emph{a posteriori} correction to ATDDFT calculations. Importantly, all the necessary ingredients except for $H_{qD}$ can be obtained solely from any code with  ATDDFT capabilities. In atomic orbital codes, $H_{qD}$ can also be reconstructed from the ATDDFT calculation, since the required two-electron integrals are used in the computation of the kernel matrix elements.
In the following, we detail this using the quantum chemistry software, NWChem but the procedure can be carried out using any code that is capable of doing ATDDFT and outputting the quantities needed to construct the matrix in Eq.~(\ref{eq:casida_matrix}) (or (\ref{eq:alternate_casida})) and the chosen dressing. A python interface with NWChem can be found at the github referenced here~\footnote{https://github.com/Dawood234/DTDDFT}.

The procedure begins with a standard ATDDFT calculation to obtain the excitation energies of the states of interest. For the states that require correction due to missing contributions from doubly-excited KS excitations, we then  extract the matrices $A$ and $B$ in the relevant truncated subspace after running the ATDDFT calculation. For instance, suppose the KS excitations labeled as $q: i \rightarrow a$ and $q': j \rightarrow b$ dominate the excitation under consideration. In that case, we extract the matrix elements $A_{qq'} = A_{ia,jb}$ and $B_{qq'} = B_{ia,jb}$ within the reduced space defined by these excitations. In NWChem these matrices are written to the output file by setting the print level to \texttt{debug}.

For computational efficiency  mentioned above, 
we will utilize the dressing corrections of Eq.~(\ref{dressing_s}) and Eq~(\ref{dressing_a}). For these, we need to obtain the following  additional quantities:
\begin{enumerate} \item KS frequencies of single and double excitations that contribute to the state of interest: The frequencies are computed from the differences in molecular orbital energies, which are readily available from the output of the ATDDFT calculation in NWChem. Specifically, the KS excitation energies for single excitations, $\nu_{q}$, are given by the energy-difference of the unoccupied and occupied orbitals involved in the transition. For the double excitation frequency $\nu_D$, we pick the orbitals for which the sum of KS frequencies lies in the vicinity of the single excitations that contribute to a given state (denoted $s_1$ and $s_2$ earlier). Additionally, for the DTDDFT$\a$ variant, the ATDDFT frequencies $\Omega_{Q}^A$,~$\Omega_{s1}^A$ and $\Omega_{s2}^A$ are also required which are typically contained in the output.

\item Two-electron integrals: The two-electron integrals are essential for computing the Hamiltonian matrix elements $H_{qD}$ between KS states. These integrals account for the coupling between different excitations in the system. In NWChem these can be obtained by appending to the end of the ATDDFT input script an \texttt{fcidump} block together with the keyword \texttt{orbitals molecular}. This will output the two-electron molecular orbital integrals in a separate file.
\end{enumerate}
Finally, we construct the dressed Casida matrix as given in Eq.~(\ref{eq:alternate_casida}) in the truncated subspace of dominant single excitations and the double excitation that couples with them. The frequency-dependence of this (small) matrix means it must be diagonalized in a self-consistent manner to obtain the corrected excitation energies and transition properties. We found, for our cases, that the frequencies were converged up to $\approx 0.02$ meV within about 5 iterations. Before determining the oscillator strengths from the resulting eigenvectors, they need to be normalized according to Eq.~(\ref{normalization}). 
\section{Example: Butadiene}
We now apply this method to compute the two lowest-lying singlet excited potential energy surfaces of butadiene, 1$^1$Bu and 2$^1$Ag, along a particular one-dimensional cut. This cut is chosen as that in Ref.~\cite{PSLPF21}, parameterized by different bond-length alternations (BLA), and along which the true surfaces cross each other, so there is a switch of energy order on either side. As discussed in the introduction, butadiene has required computationally demanding methods to accurately reproduce experimental results due to the partial double-excitation character of the 2A$_g$ state. While the 1$^1$Bu state is characterized by a dominant one-electron HOMO $\rightarrow$ LUMO transition with ionic character at the ground-state equilibrium, the 2$^1$Ag state involves substantial mixing of the HOMO$-1 \rightarrow$ LUMO, HOMO $\rightarrow$ LUMO$+1$, and two-electron excitation HOMO$^2 \rightarrow$ LUMO$^2$ configurations, and has a covalent character at the ground-state equilibrium geometry. Thus, while ATDDFT  does a reasonably good job  describing 1$^1$Bu state (even at the GGA level, the shape is well-captured), its lack of double excitation accounting makes it inappropriate and inaccurate for the  2$^1$Ag state.

The geometries of Ref.~\cite{PSLPF21} were generated across a range of BLA coordinates and are listed explicitly in the supplementary material of that work. The BLA is defined by subtracting the average double-bond length from the average single-bond length. We will use the results from the $\delta$-CR-EOMCC(2,3) method presented in that work as our reference (``exact").  
We performed a standard ATDDFT with functional PBE0 in the basis set cc-pVTZ  for these geometries to extract the excitation frequencies of 1$^1$Bu and 2$^1$Ag. For our DTDDFT, we then followed the procedure  described above using the NWChem code~\cite{nwchem}, with the truncated subspace being spanned by $q =$  HOMO$-1 \rightarrow$ LUMO, and $q' = $ HOMO $\rightarrow$ LUMO$+1$, and the double excitation $D$ as HOMO$^2 \rightarrow$ LUMO$^2$. We tested both the DTDDFT$\a$ and DTDDFT$\s$ variants in these calculations.
The iterative diagonalization is performed in the $2 \times 2$ space spanned by these excitations and convergence was reached in five iterations.  

\begin{figure}
   \centering
\includegraphics[width=1\linewidth]{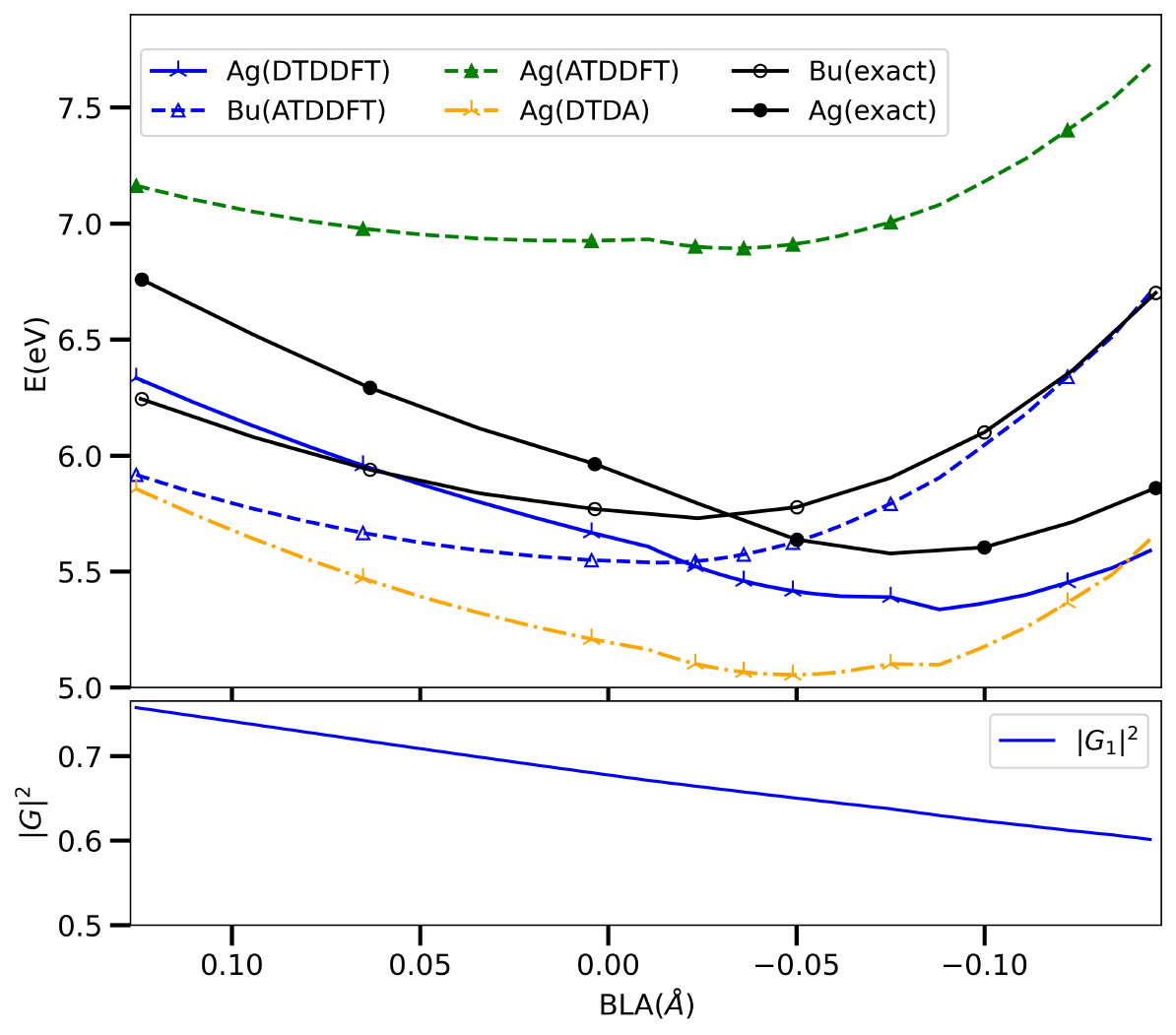}
`    \caption{Top panel: 1$^1$Bu (empty marker) and 2$^1$Ag (filled marker) excitation energies along the BLA coordinate of trans-butadiene from ATDDFT( 1$^1$Bu: blue dashed and 2$^1$Ag: green dashed) within PBE0/cc-pVTZ, ``exact" reference $\delta$-CR-EOMCC(2,3) (black), and our DTDDFT$\a$ (blue crosses) curve, and DTDA (orange dash-dot).  Bottom panel: Norm-square of the eigenvector corresponding to 2$^1$Ag of Eq.~(\ref{eq:alternate_casida}) normalized according to Eq.~(\ref{normalization}).}
    \label{fig:crossing}
\end{figure}

In the top panel of  Fig.~\ref{fig:crossing} we show the potential energy surfaces, defined as $E_n(\dulR) - E_0(\dulR_0)$ where $E_n(\dulR)$ is the energy of the $n$th excited state at the nuclear geometry $\dulR$, and $\dulR_0$ is the geometry at the Frank-Condon point. The ATDDFT energy of the 1$^1$Bu state approximates the reference $\delta$-CR-EOMCC(2,3) closely, both in magnitude and trend, but ATDDFT fails spectacularly for the 2$^1$Ag energy and completely misses the curve crossing. In contrast, the DTDDFT$\a$ results show a much closer agreement with the reference, significantly lowering the energy predicted by ATDDFT, and correcting the trend as a function of BLA coordinate. The conical intersection is predicted to be at BLA of -0.020~$\mathring{A}$ to be compared with -0.032~$\mathring{A}$ of the reference. 
 
We note that the earlier DTDA of Ref.~\cite{MZCB04, CZMB04} was applied to butadiene in Ref.~\cite{CZMB04,MW09,HIRC11,MMWA11}, also with the PBE0 functional (but in smaller basis sets). 
While single-point calculations were performed at the ground-state geometry and at the excited state minimum, the shape of the potential energy surfaces was not studied. Interestingly, we find that DTDA with PBE0/cc-pVTZ significantly underestimates the energy of the 2Ag state throughout this range of BLA, and fails to capture the curve-crossing with 1Bu. 
While this is true for any of the three variants, Fig.~\ref{fig:crossing} shows just the DTDA$\a$ variant. While we do not expect DTDA to provide accurate oscillator strengths of this state, calculations on systems studied previously~\cite{DM23} showed that DTDA performed similarly to DTDDFT for the energies themselves. 
The larger error in DTDA compared with DTDDFT could be related to the fact that the 1Bu state is predicted by PBE0/TDA about 0.5--0.9eV higher over this geometry range than PBE0/TDDFT. This error translates into an error in the denominator in the dressing of Eq. (19), leading to the error in the DTDA prediction of the 2Ag state.
A closer analysis of this is left for future work.

The bottom panel of Fig.~\ref{fig:crossing} shows the norm-square of the eigenvector of Eq.~(\ref{eq:alternate_casida}), normalized according to Eq.~(\ref{normalization}), for the  2$^1$Ag state. As argued in Ref.~\cite{DM23}, this is an estimate of the proportion of the single excitation contribution to the oscillator strength. While we do not have an exact reference for this, we note that the value around the ground-state equilibrium (BLA around 0.13$\mathring{A}$) is consistent with the  $\% T_1$ value of (74-76 \%) from the CC3 calculation quoted in Ref.~\cite{LBSCJ19}. Thus, not only the excitation energy but also the character of the state, is accurately captured by DTDDFT.

\begin{figure}
    \centering
    \includegraphics[width=1\linewidth ]{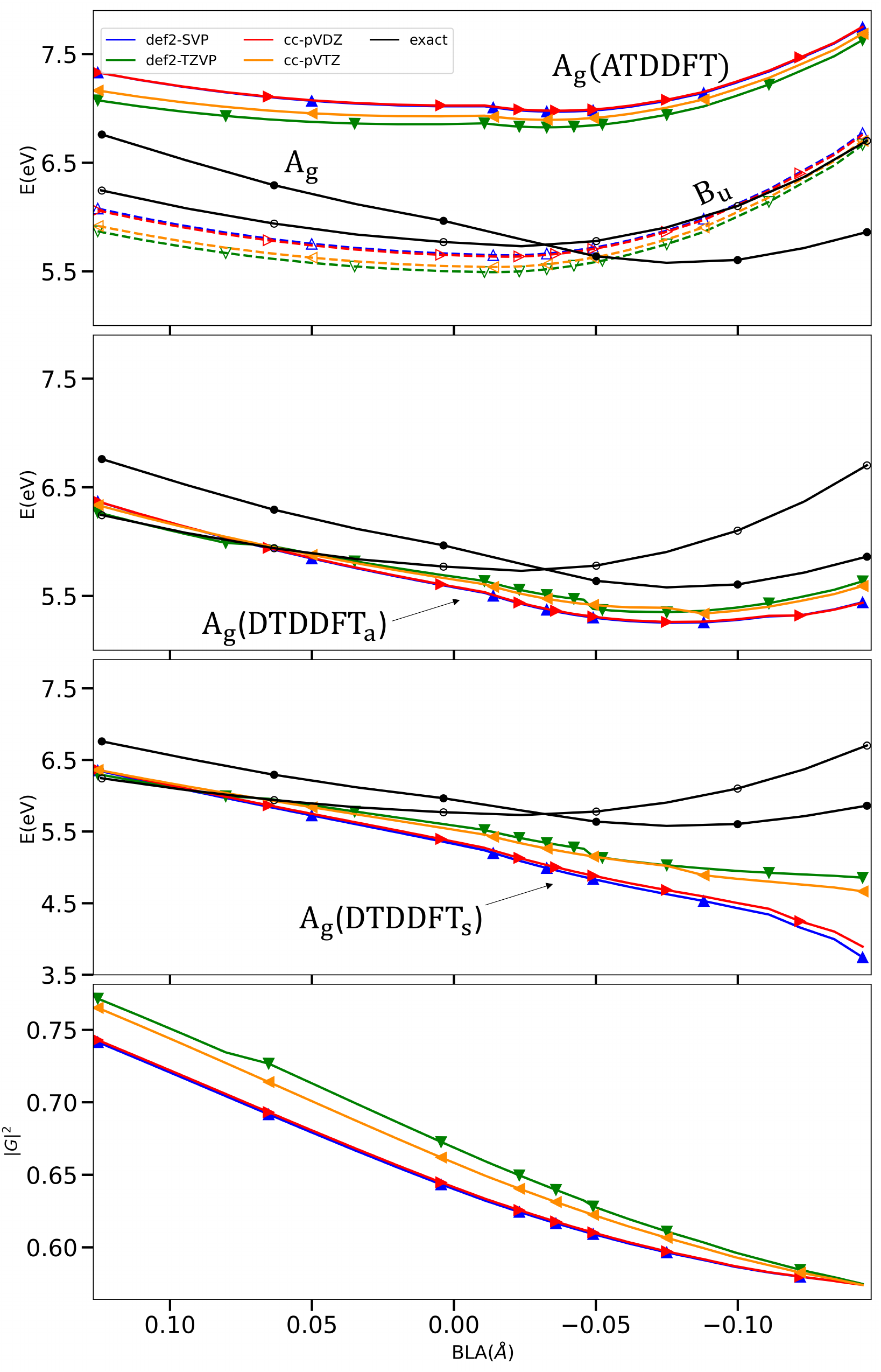}
    \caption{Energies and norm-squares of $G_I$ in basis sets def2-SVP(blue), def2-TZVP(green), cc-pVDZ(red) and cc-pVTZ(orange). Top panel: ATDDFT energies of 1$^1$Bu (empty markers) and 2$^1$Ag (filled markers) along with the corresponding reference values in black. Second panel: DTDDFT$\a$ 2$^1$Ag (filled markers). Third panel: DTDDFT$\a$ 2$^1$Ag (filled markers). Bottom panel: Norm-square, $|G_I|$ of the eigenvector corresponding to 2$^1$Ag. }
    \label{fig:basis_sets}
\end{figure}

A key consideration in the DTDDFT computation is the basis-set dependence: the $H_{qD}$ ingredient may potentially increase the basis-set dependence above what ATDDFT typically enjoys. To investigate this, the top panel of  Fig.~\ref{fig:basis_sets} shows the ATDDFT (PBE0) energies of 1$^1$Bu and 2$^1$Ag with the basis sets def2-SVP, def2-TZVP, cc-pVDZ, and cc-pVTZ. The results show little difference for either the 1$^1$Bu or 2$^1$Ag states across different BLA values. In the second panel, we examine how this behavior changes when replacing the ATDDFT 2$^1$Ag with its DTDDFT$\a$ counterpart. Notably, DTDDFT$\a$ maintains a comparable level of basis-set dependence to ATDDFT, suggesting robust numerical stability. 
The third panel shows these same basis sets for the DTDDFT$\s$ variant. First, we notice that although DTDDFT$\s$ would also capture the curve-crossing, it is less accurate than DTDDFT$\a$, and  the trend particularly deviates from the reference towards the negative BLA end of the curve. 
The figure indicates that this is partly a basis-set non-convergence issue since the results show an increasing sensitivity as the BLA goes from positive values to negative ones. The cc-pVTZ results are likely not quite converged for these larger negative values of the BLA. 
It is worth noting that the difference between different high-level wavefunction methods also increases as the BLA changes from positive to negative~\cite{PSLPF21}. We leave to future work to understand better the underlying cause of this increased sensitivity and deviation of the DTDDFT$\s$ variant at these BLAs. Why the DTDDFT$\a$ variant is more accurate than DTDDFT$\s$ may be understood from the fact that the former accounts for couplings between the KS single excitations in the dressing itself, unlike in DTDDFT$\s$, which may be important when the single-excitations couple strongly. We further note that DSMA$\a$ outperformed DSMA$\s$ for the lower excitation of the mixed single-double pair in the systems shown in Ref.~\cite{DM23}.  
Finally, we show the $|G|^2$ in the above basis sets in the lowest
panel. We observe a relatively small basis set dependence especially
for negative BLA values; in contrast to the energies, the predicted
percentage of single excitations is less sensitive at the negative BLA
end than the positive.

\subsection{Summary}
In conclusion, DTDDFT is an accurate and computationally efficient method for modeling the challenging state of double-excitation character in butadiene, yielding energies much closer to high-level wavefunction reference $\delta-$CR-EOMCC(2,3) of Ref.~\cite{PSLPF21}, and better than many other wavefunction methods~\cite{PSLPF21} that much more computationally expensive. The shape of the predicted 2Ag surface through a cut of varying BLA tracks the reference well, and
displays the elusive curve-crossing with the 1Bu state at a bond-length close to that of the reference, showing the promise of using DTDDFT for non-adiabatic dynamics calculations of interconversion processes. Whether the entire topology of the conical intersection is well-reproduced will be studied in future work.  The underlying character of the state is captured well, as further evidenced by the predicted percentage of single-excitation character, being similar to that from CC3 calculations of Ref.~\cite{LBSCJ19}.  

Thus, our DTDDFT approach offers a significant advance in the low-cost computational modeling of states of double-excitation character, overcoming key challenges that both ATDDFT and many wavefunction methods suffer from, and offers a more reliable foundation for studying ultrafast dynamics in molecular systems.  
While our earlier work~\cite{DM23} had demonstrated the fundamental idea of DTDDFT, which in turn had built upon the earlier DTDA~\cite{MZCB04,CZMB04, MW09,MMWA11,HIRC11} that was unable to correctly yield oscillator strengths and transition densities, here we have shown that DTDDFT can be efficiently extended and implemented to compute  excitations involving several single excitations mixing with a double-excitation as in the case of realistic molecules such as the low-lying polyenes. 
We observe that in our example, over most of the geometries shown,  the DTDDFT error for the state of double-excitation character Ag is similar to the adiabatic TDDFT error for the single-excitation Bu, when the DTDDFT is built from the same adiabatic xc functional, suggesting that DTDDFT might be able to be used with as much confidence for these states as ATDDFT is for single excitations, although further tests would be necessary.
The computational cost is not much beyond ATDDFT, involving only two-electron calculations in a small subspace, and the implementation can be straightforwardly interfaced with standard quantum chemistry codes. This makes it a promising tool for spectra, mapping out excited state potential energy surfaces, and for computing ultrafast non-adiabatic dynamics in photo-excited molecules. How to compute analytic gradients with our functional for efficient calculations of mixed quantum-classical dynamics is an important direction for future research. 
\subsection{Acknowledgment}
We thank Piotr Piecuch and Jun Shen for providing us with the BLA data from their work and for useful conversations. We gratefully acknowledge financial support from the National Science Foundation (NSF) under Award No. CHE-2154829 and OAC-2117429, which includes support for the High-Performance Computing Cluster (Price).
\onecolumngrid
\newpage
\renewcommand{\theequation}{S.\arabic{equation}}
\subsection{Supplementary}
\subsubsection{Reduction of DTDDFT kernel to DTDA}
Here we show that the dressed frequency-dependent kernels derived in the main text reduce to the DTDA expression from references\cite{CZMB04,MZCB04,MW09} when backward transitions are neglected, as in the Tamm-Dancoff approach. We start with the dressing in Eq.~(18) of the main text but keeping the term, $H_{qD} H_{Dq'}$ in the denominator from Eq.~(10):
\bea
X_{qq'}(\omega)=\frac{H_{qD}H_{Dq'} }{4 \sqrt{\nu_{q}\nu_{q'}}}\left[ 1 + \frac{(\nu_{q}+\nu_D)(\nu_{q'}+\nu_D)}{\omega^2 - (\nu_D^2+H_{qD} H_{Dq'})} \right]\,.
\eea 
Note that the analysis below proceeds similarly for the different variants.
The Tamm-Dancoff approximation involves ignoring the backward transitions, i.e. by considering positive-frequencies far from negative-frequency roots. 
Let us expand around the positive root of the denominator,  \( \omega_+ \), where
\begin{equation}
\omega_+ = \sqrt{\nu_D^2 + H_{qD}H_{Dq'}}.
\end{equation}
Near this root, we express \( \omega \) as \( \omega_+ + \delta \), where \( \delta \) is a small deviation:
\begin{equation}
\omega = \omega_+ + \delta.
\end{equation}
Expanding the denominator in terms of \( \delta \) gives, up to order $\delta$:
\bea
(\omega_+ + \delta)^2 - (\nu_D^2 + H_{qD}H_{Dq'})=
2\omega_+\delta \,,
\eea
so that
\begin{equation}
    \frac{1}{\omega^2 - (\nu_D^2 + H_{qD}H_{Dq'})}=\frac{1/2\omega_{+}}{\omega-\omega_{+}}.
\end{equation}
Assuming the single KS excitation frequencies, $\nu_q $ and $\nu_{q'}$ lie close to each other and also to the double excitation frequency $\nu_D$, the above dressing reduces to
\begin{equation}
    X_{qq'}(\omega)=\frac{H_{qD}H_{Dq'}}{4\nu_D}\left[1+\frac{2\nu_D^2/\omega_{+}}{\left(\omega-\omega_+\right)}\right]
\end{equation}
Near a double excitation, the last term in the brackets becomes much larger than 1, so we can write the above equation as
\begin{equation}
    X_{qq'}(\omega)=\frac{H_{qD}H_{Dq'}(\nu_D/\sqrt{\nu_D^2 + H_{qD}H_{Dq'}})}{2\left(\omega-\sqrt{\nu_D^2+H_{qD}H_{Dq'}}\right)}.
\end{equation}
Recalling that $X_{qq'}$ is added to the adiabatic part of the xc kernel, we now compare this equation with the kernel given in Ref.~\cite{MW09}, which was derived from the original dressed Tamm-Dancoff approach, and using the analog of what we call here the DTDDFT$\s$ variant. The result from Ref.~\cite{MW09} is
\begin{equation}
f\xcqqp^{\text{DTDA}}(\omega)= f\xcqqp^{\text{adia}} + \frac{H_{qD}H_{Dq'}}{2(\omega - \nu_D)},
\end{equation} from which we see that the dressing term (second term in the above equation) agrees with ours 
once we replace $\nu_D^2
+ H_{qD}H_{Dq'}$ with $\nu_D^2$. As discussed in the main text of the paper, having $H_{qD}H_{Dq'}$ in the denominator of the dressing makes it ill-defined by giving it an arbitrary sign-dependence. Here we see another reason to neglect it: ignoring this term makes our kernel agree with the one given in Ref.~\cite{MW09,MZCB04,CZMB04} derived from the dressed Tamm-Dancoff approach. 

\subsubsection{Results for PBE, PBE0 and CAM-B3LYP}
\begin{figure}[h!]
    \centering
    \includegraphics[width=1\linewidth]{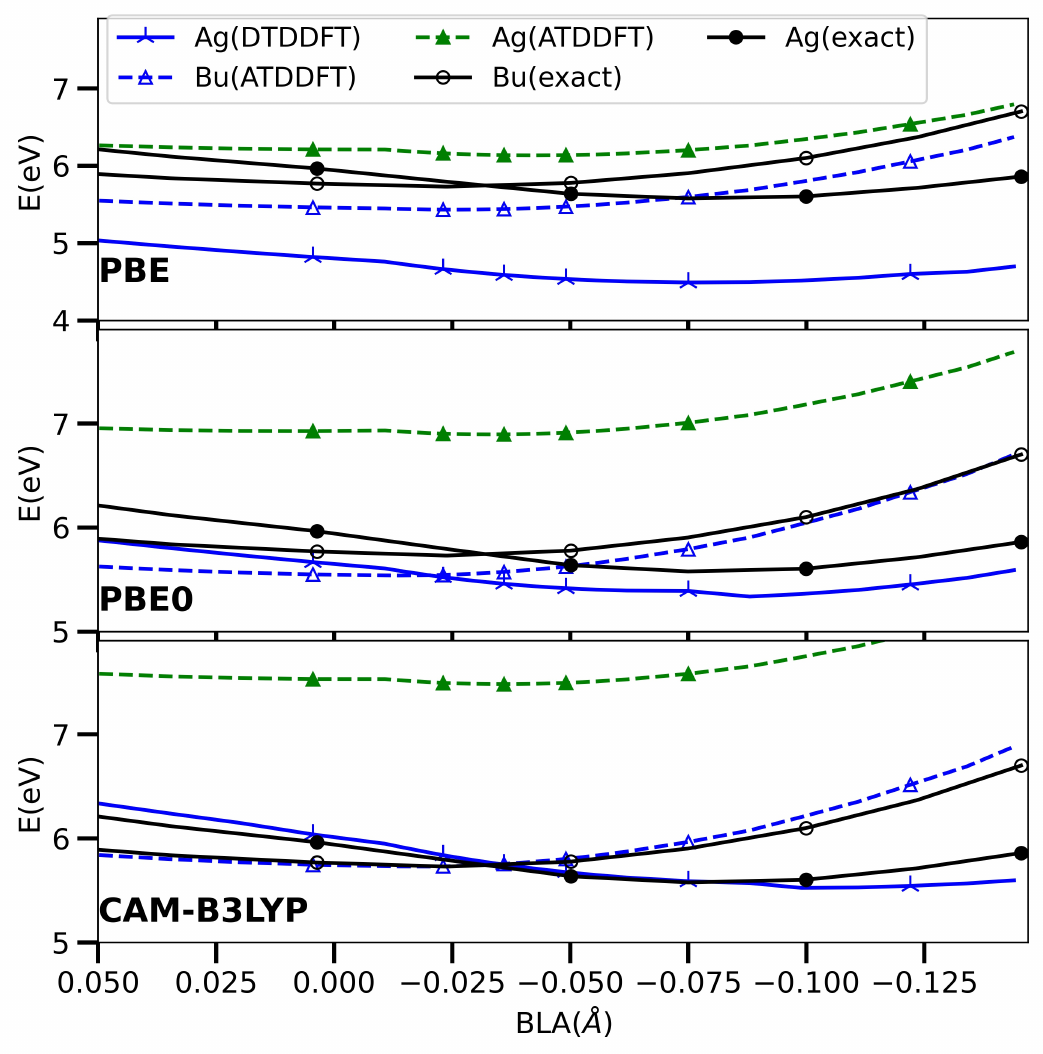}
    \caption{Comparison of DTDDFT$\a$ energies based on PBE, PBE0, and CAM-B3LYP for the adiabatic functional energies (top, middle, and lowest, respectively). As in Fig 1 of the main paper, the respective ATDDFT energies for 1Bu and 2Ag are shown, along with the reference 'exact' from $\delta$-CR-EOMCC(2,3).}
    \label{fig:func_diff}
\end{figure}
Here we investigate the performance of DTDDFT$\a$ for the butadiene curve-crossing when it is built upon three different types of adiabatic functionals: a GGA (PBE), the hybrid PBE0 that was presented in the main text, and a range-separated hybrid CAM-B3LYP. The figure shows the results computed in the def2-SVP basis. 
We see that while predictions from all three functionals  are similar for the 1Bu state, with the most accurate being CAM-B3LYP, their DTDDFT$\a$ renditions differ more for the 2Ag state, again with CAM-B3LYP being most accurate. Although the trend of the exact curve is well-captured (unlike adiabatic PBE), DTDDFT$\a$ with PBE is a significant underestimate of about 1eV, too low to display a crossing in the right region. This underestimate can be seen to be largely inherited from the underestimate of the adiabatic energy rather than the dressing per se. 
On the other hand the CAM-B3LYP is remarkably accurate. However, it should be noted that we were not able to confirm that the CAM-B3LYP result shown is converged with respect to basis set (as we do for PBE0 in the Fig. 2 of the main paper); the range-separated hybrid appears to yield more sensitivity to basis set in our DTDDFT. A deeper study of why this is so is beyond the scope of the present study.

\bibliography{main.bib}
\end{document}